\documentclass[doublecol]{epl2}

\usepackage{amsmath}
\usepackage{amssymb}
\usepackage{psfrag}

\usepackage{graphicx}

\newcommand{\av}[1]{\left\langle{#1} \right\rangle}  

\title{Global dynamics of oscillator populations under common noise}
\author{W. Braun \inst{1,2} \and  A. Pikovsky \inst{1} \and  M. A. Matias \inst{3} \and P. Colet \inst{3}}
\institute{\inst{1} Department of Physics and Astronomy, Potsdam University, 14476 Potsdam-Golm, Germany\\
\inst{2}Department of Applied Mathematics and Theoretical Physics, University of Cambridge, Centre for Mathematical Sciences, 
Cambridge CB3 0WA, UK\\
\inst{3} IFISC, CSIC-UIB, Campus UIB,
07122 Palma de Mallorca, Spain}


\abstract{ Common noise acting on a population of identical oscillators 
can synchronize them. We develop a description of this process which is 
not limited to the states close to synchrony, but provides a global 
picture of the evolution of the ensembles. The theory is based on the 
Watanabe-Strogatz transformation, allowing us to obtain closed 
stochastic equations for the global variables. We show that on 
the initial stage, the order parameter grows linearly in time, 
while on the late stages the convergence to  synchrony is 
exponentially fast. Furthermore, we extend the theory to nonidentical
ensembles 
with the Lorentzian distribution of natural frequencies and determine 
the stationary values of the order parameter in dependence on driving 
noise and mismatch.   
}
\begin{document}

\maketitle

\section{Introduction}
Synchronization of oscillations by a periodic forcing is a general 
phenomenon observed in numerous experiments. In this setup the 
system follows the driving and has, in particular, the same 
frequency, so one often speaks on frequency entrainment. Much more 
nontrivial is the effect of synchronization by an external noise. 
Here one can also distinguish between the cases when the driven 
system is entrained by the noise (synchrony) and the situations when the 
noise is not followed (asynchrony). While the difference between 
these regimes  can be hardly seen by observing just one responding 
oscillator, it becomes evident if an ensemble of identical systems 
driven by the same noise is observed: in the case of synchronization 
all the oscillators in the ensemble follow the forcing and their 
states thus coincide, while in the asynchronous state the states of 
systems remain different. This effect is therefore called synchronization 
by common 
noise~\cite{Pikovsky-84,Pikovsky-84a,Pikovsky-Rosenblum-Kurths-01,%
Goldobin-Pikovsky-04,Teramae-Tanaka-04,Goldobin-Pikovsky-05a}. 
An interesting realization of this type of synchronization is 
the effect of reliability of neurons~\cite{Mainen-Sejnowski-95}. 
Here one does not use an ensemble of identical neurons, but takes 
one neuron and applies the same pre-recorded noise to it several times. 
The synchronous case appears as a reliable respond to the forcing where 
all the noise-induced spikes are at the same
 position at all runs, while for asynchrony (antireliability) 
 the same noise produces different spike 
 patterns~\cite{Goldobin-Pikovsky-05b,Goldobin-Pikovsky-06b}. 
 Synchronization by common noise was also observed in physical experiments 
 with phase-locked loop~\cite{Khoury-Lieberman-Lichtenberg-98} 
 and noise-driven lasers~\cite{Uchida-Mcallister-Roy-04}.  
 
 Synchronization by common noise can be characterized by the largest 
 Lyapunov exponent of the noise-driven dynamics. This exponent 
 governs the growth/decay  of small perturbations of a synchronous 
 state; a negative exponent corresponds to synchrony while a positive 
 one to asynchrony~\cite{Pikovsky-84,Pikovsky-84a,Pikovsky-Rosenblum-Kurths-01,%
Goldobin-Pikovsky-04,Teramae-Tanaka-04,Goldobin-Pikovsky-05a} (notice that here the largest Lyapunov exponent can be interpreted as a ``transverse'' one, determining the growth/decay of the difference between oscillators in the ensemble). For periodic oscillators, which in the noise-free 
 case have a zero maximal Lyapunov exponent, small noise generally 
 leads to a negative exponent (while large noise can desynchronize); 
 for chaotic systems with a positive Lyapunov exponent, strong noise 
 can synchronize (see examples of the synchronization-desynchronization 
 transition in~\cite{Pikovsky-84,Pikovsky-84a,Yu-Ott-Chen-90,Pikovsky-92c,Toral01}). 
 Calculation of the Lyapunov exponent is a relatively easy 
 numerical task, and in many cases it can 
 be obtained 
 analytically~\cite{Goldobin-Pikovsky-04,Teramae-Tanaka-04,%
 Goldobin-Pikovsky-05a}.  
 This theory is, however, restricted to the linear analysis of a stability of the synchronous state, and does not allow one to follow the evolution starting from a broad distribution of the phases.
 
 The goal of this letter is to present a \textit{global} analytic theory of the synchronization by common noise, i.e. a theory describing the evolution 
 toward synchrony of the population starting from the distributed, asynchronous state. Our theory is based on the Watanabe-Strogatz ansatz~\cite{Watanabe-Strogatz-94} that allows one to describe a population of phase oscillators under common forcing via closed equations for the macroscopic, global variables. We will show that the resulting equations can be written as a noise-driven Hamiltonian system, and will 
 analyze the evolution of ensembles close to synchrony (where the results of the Lyapunov analysis will be recovered) as well as the evolution starting from the maximally asynchronous state.

\section{Global variables description of ensembles}

Our goal is to describe an ensemble of identical (phase) oscillators,
and one can show that the following model is general enough to describe all
interesting cases:
\begin{equation}
\dot\varphi_k=\Omega(t)+\text{Im}(F(t)e^{-i\varphi_k}),\quad k=1,\ldots,N\;.
\label{eq:basphi}
\end{equation}
Here  oscillators are described by their phases $\varphi_k$, and  $\Omega(t)$ and $F(t)$ are time-dependent common forces (without loss of generality we can assume that $F(t)$ is real, otherwise one shifts the phases $\varphi$ by the argument of complex $F$ and correspondingly redefines $\Omega$). 
We want to characterize the evolution of the ensemble especially for noisy forces.

Now we present two particularly important applications, which can be described using Eqs.~(\ref{eq:basphi}). One is an ensemble of self-sustained oscillators under common external force. Unforced oscillators are described by $\dot\varphi_k=\omega$, and a noisy forcing is typically described by 
\begin{equation}
\dot\varphi_k=\omega-\sigma \xi(t) \sin \varphi_k\;.
\label{eq:baso}
\end{equation}
This system, previously considered in \cite{Goldobin-Pikovsky-04,Goldobin-Pikovsky-05a}, reduces to Eq.~(\ref{eq:basphi}) with constant $\Omega=\omega$ and 
 $F(t)=\sigma\xi(t)$. Another relevant physical setup 
is a sequential array of shunted Josephson junctions subject to a common current $I(t)$:
\begin{equation}
\frac{\hbar}{2eR} \frac{d\varphi_k}{dt}+I_c\sin \varphi_k=I(t)\;.
\label{eq:basj0}
\end{equation}
Here $\varphi_k$ is the junction's phase (difference of the phases of the
macroscopic wave functions in superconductors constituting the junction), $R$ is the resistance of the shunt, $I_c$ is the critical current. Supposing that the current $I(t)$ has a constant and  noisy component 
$I=I_0+I_1(t)$,then by rescaling time $t\to \frac{I_c2eR}{\hbar}t$,
we can write system (\ref{eq:basj0}) as 
\begin{equation}
\dot\varphi_k=\omega+\sigma\xi(t)-\sin \varphi_k\;,
\label{eq:basj}
\end{equation}
which is also Eq.~(\ref{eq:basphi}) with constant 
$F=1$ and time-dependent $\Omega(t)=(I_0+I_1(t))(I_c)^{-1}=\omega+\sigma\xi(t)$.

In the seminal work~\cite{Watanabe-Strogatz-94} 
Watanabe and Strogatz (WS)  demonstrated that the ensemble (\ref{eq:basphi}), for any $N>3$, can be fully described with three global variables and $N-3$ constants of motion. We will use here the formulation of the WS theory given in~\cite{Pikovsky-Rosenblum-11}. The transformation to the global variables $\rho,\Phi,\Psi$ and constants $\psi_k$ is performed according to
\begin{equation}
e^{i\varphi_k}=e^{i\Phi}\frac{\rho+e^{i(\psi_k-\Psi)}}{\rho e^{i(\psi_k-\Psi)}+1}\;,
\label{eq:wsdef}
\end{equation}
with an additional condition $\sum_k e^{i\psi_k}=0$. 
The closed system of equations for $\rho,\Phi$ 
reads 
(as $\Psi$ does not enter in the dynamical equations for
$\dot{\rho}$ and $\dot{\Phi}$, it does not to be taken into account.)
\begin{equation}
\begin{aligned}
\dot\rho&=\frac{1-\rho^2}{2}\text{Re}(F(t)e^{-i\Phi})\;,\\
\dot\Phi&=\Omega(t)+\frac{1+\rho^2}{2\rho}\text{Im}(F(t)e^{-i\Phi})\;.
\end{aligned}
\label{eq:ws1}
\end{equation}

The physical meaning of the global variables $\rho,\Phi$ is clear from their definition  (\ref{eq:wsdef}). 
The case of uniformly spread
constants of motion $\psi_k$ is easiest to interpret, because
as it has been shown in~\cite{Pikovsky-Rosenblum-08,Pikovsky-Rosenblum-11}, for a uniform distribution of constants $\psi_k$ one has $\rho\exp(i\Phi)=N^{-1}\sum_k \exp(i\varphi_k)$. This means that $z=\rho e^{i\Phi}$ is the complex Kuramoto order parameter widely used for characterizing synchrony in the ensemble. Thus, $\rho$ is roughly proportional to the mean field amplitude:
for $\rho=0$ the phases $\varphi_k$ are uniformly spread while 
for $\rho=1$ they form a cluster (from which at most one oscillator 
with $\psi_k-\Psi=\pi$ may deviate) of perfect synchrony. The variable $\Phi$, being the phase of the mean field, characterizes the position of the maximum in the distribution of phases. Finally, the variable $\Psi$ characterizes the offset of the phases of individual oscillators with respect to $\Phi$. For a non-uniform distribution of the constants  $\psi_k$,  the complex variable $z$ does not coincide with the Kuramoto order parameter; nevertheless, the limits $\rho\to 0$ and $\rho\to 1$ correspond to fully asynchronous and fully synchronous cases, so $\rho$ yields a proper characterization of synchrony. 

\section{Hamiltonian formulation}
Our goal in this paper is to describe the evolution of the  ensemble (\ref{eq:basphi}) for noisy $\Omega(t),F(t)$, by virtue of the global variables dynamics  (\ref{eq:ws1}). Remarkably, one can reformulate the basic equations  (\ref{eq:ws1}) as a \textit{Hamiltonian} system~\cite{Braun-BA}. Indeed, in variables 
\begin{equation}
q=\frac{\rho\cos\Phi}{\sqrt{1-\rho^2}}\;,\qquad p=-\frac{\rho\sin\Phi}{\sqrt{1-\rho^2}}\;,
\label{eq:tran}
\end{equation}
the equations read
\begin{equation}
\begin{aligned}
\dot q&=\frac{\partial H}{\partial p}=\Omega(t) p+F(t)\frac{1+q^2+2p^2}{2\sqrt{1+p^2+q^2}}\;,\\
\dot p&=-\frac{\partial H}{\partial q}=-\Omega(t) q-F(t)\frac{qp}{2\sqrt{1+p^2+q^2}}\;,
\end{aligned}
\label{eq:pqeq}
\end{equation}
with Hamiltonian 
\begin{equation}
H(q,p,t)=\Omega(t)\frac{p^2+q^2}{2}+F(t)\frac{p\sqrt{1+p^2+q^2}}{2}\;.
\label{eq:ham-qp}
\end{equation}
 One can also formulate the dynamics in ``action-angle'' variables~\cite{Goldstein-59}, where the angle is the WS variable $\Phi$ and the action is defined according to 
\begin{equation}
J=\frac{q^2+p^2}{2}=\frac{\rho^2}{2(1-\rho^2)}\;.
\label{eq:acan}
\end{equation}
The Hamiltonian equations now read
\begin{equation}
\begin{aligned}
\dot J&=F(t)\frac{\sqrt{2J(2J+1)}}{2}\cos(\Phi)=-\frac{\partial H}{\partial \Phi}\;,\\
\dot\Phi&=\Omega(t)-F(t)\frac{4J+1}{2\sqrt{2J(2J+1)}}\sin(\Phi)=\frac{\partial H}{\partial J}\;,
\end{aligned}
\label{eq:hamaa}
\end{equation}
with Hamiltonian, 
\begin{equation}
H(J,\phi,t)=\Omega(t) J-F(t)\frac{\sqrt{2J(2J+1)}}{2}\sin\Phi\;.
\label{eq:ham-aa}
\end{equation}
The action $J$ yields, according to its relation to the Kuramoto order parameter $\rho$, a natural characterization of synchrony in the population of oscillators: $J\to 0$ corresponds to a maximally asynchronous, uniformly spread state, while $J\to\infty$ corresponds to a perfect synchrony where all oscillators cluster in a single point in phase space.

The Hamiltonian formulation allows us to give a general qualitative
description of the dynamics. Notice that since the Hamiltonian is time-dependent, the energy in eqs.~(\ref{eq:pqeq},\ref{eq:hamaa}) is not conserved. Typically, noise leads to a growth of energy, either diffusive or exponential (in exceptional cases, e.g. if in (\ref{eq:ham-aa}) $F(t)$ vanishes, the system possesses an integral and no growth of energy is observed). Thus, the action variable $J$ grows and the system tends to synchrony. We analyze below this growth for large and small values of $J$, i.e. close to synchrony and close to asynchrony, using $\av{J}$ as the order parameter characterizing the level of synchrony.

\section{Dynamics close to synchrony}
Close to synchrony, i.e. for $J\gg 1$, we can approximate the Hamiltonian (\ref{eq:ham-aa}) as $H(J,\phi,t)=(\Omega(t)-F(t)\sin\Phi)J$ which leads to a skew system where the dynamics of $\Phi$ does not depend on $J$:
 \begin{align}
\dot J&=F(t)J\cos(\Phi)\label{eq:ham-lin1}\;,
\\
\dot\Phi&=\Omega(t)-F(t)\sin(\Phi)\;.
\label{eq:ham-lin2}
\end{align}
This yields $\ln J(t)=\ln J(0)+\int_0^t F(t')\cos\Phi(t')dt'$ and $\Phi(t)$ is a solution of (\ref{eq:ham-lin2}). On average, $\ln J$ grows linearly in time with the rate given by the  Lyapunov exponent
\begin{equation}
\av{\frac{d}{dt}\ln J}=-\lambda=\av{F(t)\cos(\Phi)}\;.
\label{eq:lex}
\end{equation}
The same Lyapunov exponent appears when one directly analyses stability of the cluster $\varphi_1=\varphi_2=\ldots=\varphi_N=\Phi$ in system (\ref{eq:basphi}). The phase of the cluster obeys (\ref{eq:ham-lin2}) and the small deviation of one of the phases $\delta\varphi$ obeys 
$$
\frac{d}{dt}\delta\varphi=-F(t)\cos(\Phi)\delta\varphi\;,
$$
what means that $\delta\varphi\sim\exp(\lambda t)$ with the Lyapunov exponent defined in (\ref{eq:lex}). For both applications (\ref{eq:baso}) and (\ref{eq:basj}), the calculations of the Lyapunov exponent for a white Gaussian noise $\xi(t)$ have been already reported in the literature (see Fig. 9.4 in book~\cite{Pikovsky-Rosenblum-Kurths-01} for (\ref{eq:basj}) and refs.~\cite{Goldobin-Pikovsky-04,Teramae-Tanaka-04,Goldobin-Pikovsky-05a} for (\ref{eq:baso})). 

\section{Dynamics close to asynchrony}
Here we describe the dynamics for small values of $J$, i.e. close to the asynchronous regime with nearly uniform distribution of the phases $\varphi_k$. We perform the analysis separately for the noise-driven oscillators (\ref{eq:baso}) and noise-driven Josephson junctions (\ref{eq:basj}).

\begin{figure}
\centering
\includegraphics[width=\columnwidth]{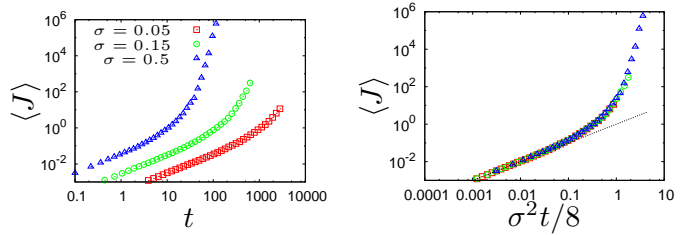}
\caption{(Color online) Evolution of $\av{J(t)}$ for noise-driven oscillators for different $\sigma$ (right panel: the same in a rescaled time), for $\omega=2$ and different noise intensities $\sigma$. Dashed line corresponds to relation (\ref{eq:sol_lin}).}
\label{fig:gr-osc}
\end{figure}

An ensemble of noise-driven oscillators (\ref{eq:baso}) is described by Hamiltonian  (\ref{eq:ham-qp}) 
with $F(t)=\sigma\xi(t)$ and $\Omega=\omega=const$. For small $q,p\ll 1$ 
we can approximate the Hamiltonian as, 
\begin{equation}
H(q,p,t)=\omega\frac{p^2+q^2}{2}+\sigma\xi(t)\frac{p}{2}\;.
\label{eq:ham_soo}
\end{equation}
Assuming for simplicity of presentation that we start from the vanishing order parameter, i.e. $q(0)=p(0)=0$, we can easily solve the resulting linear equations:
\begin{equation}
q(t)+ip(t)=\frac{\sigma}{2}\int_0^t\exp[i\omega(t'-t)]\xi(t')dt',
\label{eq:sol_soo}
\end{equation}
which yields for the action $J$, after averaging,
$$
\begin{array}{l}
\av{J(t)}=\frac{1}{2}\av{q^2+p^2}=\\[1em]
=\frac{\sigma^2}{8}\int_{0}^{t} 
\int_{0}^{t} e^{i\omega(t'-t'')} 
\av{\xi(t')\xi(t'')} dt' dt''
\end{array} 
$$
Introducing $\tau=t'-t''$ and assuming that $\av{\xi(t')\xi(t'')}$ depends only on $\tau$ one can 
integrate over $t'$ so that
\begin{equation}
\begin{aligned}
\av{J(t)}=t\frac{\sigma^2}{8}\int_{-t}^t\av{\xi(0)\xi(\tau)}\cos\omega \tau (1-\frac{|\tau|}{t})d\tau\;.
\end{aligned}
\label{eq:sol_soo_j}
\end{equation}
Asymptotically, for large $t$ this describes a diffusive linear growth of $J$
(provided the integral converges, i.e. the correlation function of noise decays fast enough); in the case of white noise, when $ \av{\xi(0)\xi(t')}=\delta(t')$, we get 
\begin{equation}
\av{J(t)}=\frac{\sigma^2}{8}t\;. 
\label{eq:sol_lin}
\end{equation}
In Fig.~\ref{fig:gr-osc} we compare for model (\ref{eq:baso}) the theoretical results with the numerical ones (obtained from eqs.~(\ref{eq:pqeq}),(\ref{eq:hamaa})).

\begin{figure}
\centering
\includegraphics[width=\columnwidth]{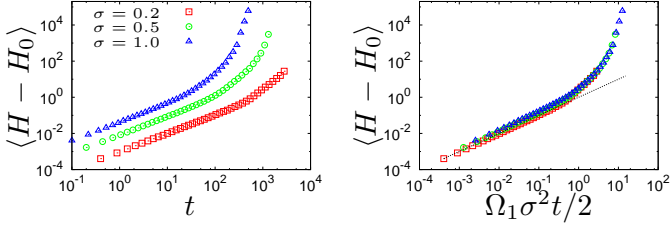}
\caption{(Color online) Evolution of $\av{H-H_0}$ for the ensemble of Josephson junctions (\ref{eq:basj})  for $\omega=5$ and different $\sigma$ (right panel: in a rescaled time). The dashed line corresponds to formula (\ref{eq:ham_soj_ling}).}
\label{fig:gr-jj}
\end{figure}

In the case of noise-driven Josephson junctions (\ref{eq:basj}) we need to analyze the Hamiltonian
\begin{equation}
H(J,\Phi,t)=(\omega+\sigma\xi(t))J-\frac{\sqrt{2J(2J+1)}}{2}\sin\Phi\;.
\label{eq:ham_soj}
\end{equation}
We restrict our attention to the nontrivial case $\omega>1$, so that the Hamiltonian is bounded from below.
The ``lowest-energy'' state here for $\sigma=0$
is the steady state with $\Phi_0=\pi/2$, $J_0=(\omega-\sqrt{\omega^2-1})/(4\sqrt{\omega^2-1})$.
This steady state does not correspond to full asynchrony (which is not a stationary solution of the equations, because the phases rotate non-uniformly), but to a stationary distribution of the phases of the ensemble (\ref{eq:basj}). Close to this equilibrium, we can linearize the equations of motion, so that the Hamiltonian in the vicinity of $J_0,\Phi_0$ reads
\begin{equation}
H(\delta J,\delta\Phi,t)=H_0+\sigma\xi(t)\delta J+\frac{\omega_1(\delta\Phi)^2+\omega_2(\delta J)^2}{2}\;,
\label{eq:ham_soj_lin}
\end{equation}
with
$
\omega_1=0.5\sqrt{2J_0(2J_0+1)}$,  $\omega_2=2(\omega^2-1)(2J_0(2J_0+1))^{-1/2}$.
The solution of the linear equations of motion (starting from the equilibrium point)
is, similar to (\ref{eq:sol_soo}), 
$$
\delta\Phi+i\frac{\omega_2}{\kappa}\delta J=\sigma\int_0^t \exp[i\kappa(t'-t)]\xi(t')dt'\;,
$$
where $\kappa^2=\omega^2-1$.
After averaging we obtain
$$
\begin{gathered}
\av{(\delta\Phi)^2+\frac{\omega_2^2}{\kappa^2}(\delta J)^2}=\\=t\sigma^2\int_{-t}^t\av{\xi(0)\xi(\tau)}\cos\kappa \tau(1-\frac{|\tau|}{t})d\tau\;.
\end{gathered}
$$
This relation means that, asymptotically, the ``energy'' defined as $H-H_0=0.5(\omega_1(\delta\Phi)^2+\omega_2(\delta J)^2)$ grows  linearly in time, for the white noise we get
\begin{equation}
\av{H-H_0}=\frac{\omega_1\sigma^2}{2}t\;.
\label{eq:ham_soj_ling}
\end{equation}
This relation is checked in in Fig.~\ref{fig:gr-jj}, where numerical simulations of eqs.~(\ref{eq:pqeq}),(\ref{eq:hamaa}) are presented.

The results above show that there are two stages for the transition to synchrony in ensembles of oscillators driven by common noise. As the system is described by a noise-driven Hamiltonian, it is natural to characterize the evolution through the ``energy'' -- the value of the noise-independent part of the Hamilton function. At the initial stage, close to asynchrony, the growth of the energy is diffusive, its averaged square grows linearly in time according to (\ref{eq:sol_lin}) and (\ref{eq:ham_soj_ling}). When the energy reaches a 
level of order one, a crossover to the other type of behavior, namely to an exponential growth of energy, is observed from numerical simulation (cf. 
Figs.~\ref{fig:gr-osc},\ref{fig:gr-jj}). This latter stage means that the final convergence of the ensemble of oscillators to a synchronous cluster is exponentially fast. 

\section{Nonidentical oscillations}
\begin{figure}
\centering
\includegraphics[width=0.85\columnwidth]{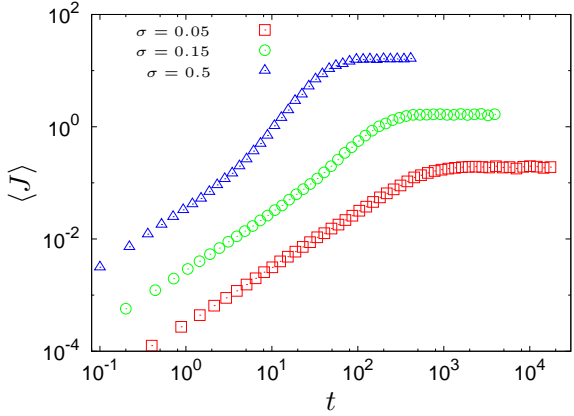}
\caption{(Color online) Evolution of the order parameter according to Eqs.~(\ref{eq:pqeq_g}),(\ref{eq:hamaa_g}) for $\omega_0=2$, $\gamma=10^{-3}$ and different noise intensities $\sigma$.}
\label{fig:ni1}
\end{figure}

Here we extend the theory to the case of nonidentical oscillators (\ref{eq:baso}) having a distribution of frequencies $g(\omega)$. In this case one first generalizes the WS description by assuming the frequency dependence of the variables $\rho(\omega),\Phi(\omega)$~\cite{Pikovsky-Rosenblum-11}. It is convenient to introduce one complex variable $z(\omega)=\rho e^{i\Phi}$ that obeys
\begin{equation}
\dot z=i\omega z+\frac{\sigma \xi(t)(1-z^2)}{2}\;.
\label{eq:zw}
\end{equation}
The distribution of the phases is now characterized by the global order parameter $Z=\int z(\omega)g(\omega)d\omega$. Following the approach of Ott and Antonsen~\cite{Ott-Antonsen-08}, it is possible to obtain a closed equation for this order parameter in the case of a Lorentzian distribution $g(\omega)=\pi^{-1}\gamma(\gamma^2+(\omega-\omega_0)^2)^{-1}$. Then, assuming that $z(\omega)$ as function of complex $\omega$ does not have singularities in the upper half-plane, one can perform the integration to get $Z=z(\omega_0+i\gamma)$. Thus, the equation for $Z$ follows from (\ref{eq:zw}) with $\omega\to\omega_0+i\gamma$:
\begin{equation}
\dot Z=i\omega_0 Z-\gamma Z+\frac{\sigma\xi(t)(1-Z^2)}{2}\;.
\label{eq:Z}
\end{equation}
Transforming from $Z$ to the canonical variables as in (\ref{eq:tran}),(\ref{eq:acan}), we obtain the same equations as (\ref{eq:pqeq}) and (\ref{eq:hamaa}) but with additional non-Hamiltonian, damping terms:
\begin{equation}
\begin{aligned}
\dot q&=\omega_0 p+\sigma \xi(t)\frac{1+q^2+2p^2}{2\sqrt{1+p^2+q^2}}-\gamma q(1+p^2+q^2)\;,\\
\dot p&=-\omega_0 q-\sigma\xi(t)\frac{qp}{2\sqrt{1+p^2+q^2}}-\gamma p(1+p^2+q^2)\;,
\end{aligned}
\label{eq:pqeq_g}
\end{equation}
and 
\begin{equation}
\begin{aligned}
\dot J&=\sigma\xi(t)\frac{\sqrt{2J(2J+1)}}{2}\cos(\Phi)-2\gamma J(1+2J)\;,\\
\dot\Phi&=\omega_0-\sigma\xi(t)\frac{4J+1}{2\sqrt{2J(2J+1)}}\sin(\Phi)\;.
\end{aligned}
\label{eq:hamaa_g}
\end{equation}
With damping terms, the energy does not grow indefinitely, but saturates as shown in Fig.~\ref{fig:ni1}. The saturation level corresponds to a bunch of oscillators that do not form a perfect cluster, but have a finite spread.

The stationary level of the order parameter can be estimated for a state close to synchrony and for a wide distribution (asynchrony). Close to synchrony, i.e. for $J\gg 1$, we have instead of (\ref{eq:lex}) 
\begin{equation}
\av{\frac{d}{dt}\ln J}=\sigma\xi(t)\cos\Phi-2\gamma J\;.
\label{eq:lex_g}
\end{equation}
If we neglect the fluctuations of the growth rate and assume $\sigma\xi(t)\cos\Phi=\av{\sigma\xi(t)\cos\Phi}=-\lambda$, then the stationary value of $J$ is $J_{st}=|\lambda|/(2\gamma)$. Close to asynchrony we can use the same approximation as in (\ref{eq:ham_soo}), but now the equations read
$$
\dot q=\omega_0 p-\gamma q+0.5\sigma \xi(t)\;,\quad \dot p=-\omega_0q-\gamma p\;,
$$
with the average stationary energy
$$
\av{q^2+p^2}=\frac{\sigma^2}{8\gamma}\int_{-\infty}^\infty\av{\xi(0)\xi(t)}\cos\omega_0 te^{-\gamma|t|}dt\;,
$$
which in the case of the white noise yields $\av{J}=\sigma^2/(16\gamma)$. 

\begin{figure}
\centering
\includegraphics[width=0.85\columnwidth]{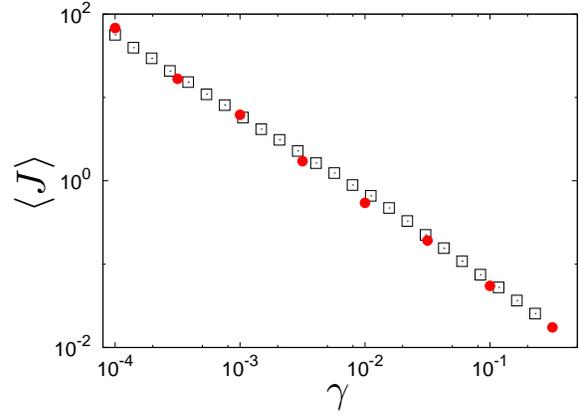}
\caption{(Color online) The stationary level of the order parameter for an ensemble of the oscillators with a Lorentzian distribution of frequencies, vs. the distribution width. Parameters: $\omega_0=2$, $\sigma=0.3$. Squares: modeling eqs.~(\ref{eq:pqeq_g}),(\ref{eq:hamaa_g}); filled circles: modeling the ensemble of $N=5000$ oscillators for the same parameters. }
\label{fig:ni2}
\end{figure}

Remarkably, in both limits  the average value of $J$ scales as $\gamma^{-1}$. This is confirmed by numerics presented in Fig.~\ref{fig:ni2}. In this figure we present also simulations of the oscillator populations, which fit nicely the results from the modeling of the WS variables.

\section{Discussion} In this letter we have developed a global theory of synchronization of oscillator populations by common noise. Our analysis 
is based on the Watanabe-Strogatz ansatz~\cite{Watanabe-Strogatz-94}, which is not restricted by a number of elements in the populations and results in an explicit time-dependence 
of the global variables on the common forcing terms. These variables can be interpreted as order parameters characterizing the population of identical oscillators. For noisy forcing we thus obtained a closed set of stochastic differential equations for the global variables. An important step in our consideration is a representation of the WS equations as a nonautonomous Hamiltonian system; transition to synchrony then appears as the growth of energy due to the noisy driving. While for the situations close to synchrony the results are essentially the same as the previous ones derived from the linear perturbation approach, we have demonstrated that when starting from a broad initial distribution, the energy first grows linearly, and only after a formation of a concentrated cluster the exponential convergence to synchrony sets on. Furthermore, by virtue of the Ott-Antonsen theory~\cite{Ott-Antonsen-08} we have extended the analysis to populations of non-identical oscillators with a Lorentzian distribution of natural frequencies. Here the theory is valid in the thermodynamic limit of very large ensembles only.

\acknowledgments
A.P. thanks IFISC for hospitality. We are very thankful to M. Rosenblum for helpful discussions. Financial support from the Spanish 
MINECO and FEDER
  under grants FISICOS (FIS2007-60327) and 
  DeCoDicA (TEC2009-14101)
  is acknowledged.


\end{document}